\documentclass[12pt,preprint2]{emulateapj}

\usepackage{natbib}
\newcommand{\beq}{\begin{equation}}
\newcommand{\beqa}{\begin{eqnarray}}
\newcommand{\eeq}{\end{equation}}
\newcommand{\eeqa}{\end{eqnarray}}

\newcommand{\simgt}{\lower.5ex\hbox{$\; \buildrel > \over \sim \;$}}
\newcommand{\simlt}{\lower.5ex\hbox{$\; \buildrel < \over \sim \;$}}

\newcommand{\bd}[1]{\mbox{\boldmath $#1$}}

\shorttitle{Mask Effect in Weak Lensing}
\shortauthors{Shirasaki et al.}

\begin{document}

\title{Effect of Mask Regions on Weak Lensing Statistics}

\author{Masato Shirasaki}
\affil{Department of Physics, University of Tokyo, Tokyo 113-0033, Japan}
\email{masato.shirasaki@utap.phys.s.u-tokyo.ac.jp}

\author{Naoki Yoshida}
\affil{
Department of Physics, University of Tokyo, Tokyo 113-0033, Japan\\
Kavli Institute for the Physics and Mathematics of the Universe (WPI),
University of Tokyo, Kashiwa, Chiba 277-8583, Japan
}

\and 

\author{Takashi Hamana}
\affil{National Astronomical Observatory of Japan, Tokyo 181-0015, Japan}

\begin{abstract}
Sky masking is unavoidable in wide-field weak lensing observations.
We study how masks affect the measurement of statistics 
of matter distribution probed by weak gravitational lensing.
We first use 1000 cosmological ray-tracing simulations to examine 
in detail the impact of masked regions on the weak lensing Minkowski Functionals (MFs).
We consider actual sky masks used for a Subaru Suprime-Cam imaging survey.
The masks increase the variance of the convergence field
and the expected values of the MFs are biased.
The bias then affects the non-Gaussian signals 
induced by the gravitational growth of structure.
We then explore how masks affect cosmological parameter estimation. 
We calculate the cumulative signal-to-noise ratio S/N for masked maps 
to study the information content of lensing MFs.
We show that the degradation of S/N for masked maps is mainly 
determined by the effective survey area.
We also perform simple $\chi^2$ analysis to show the impact of 
lensing MF bias due to masked regions.
Finally, we compare ray-tracing simulations with 
data from a Subaru 2 deg$^2$ survey
in order to address if the observed lensing MFs are consistent 
with those of the standard cosmology.
The resulting $\chi^2/n_{\rm dof} = 29.6/30$ for three combined MFs,
obtained with the mask effects taken into account,  
suggests that the observational data are indeed consistent with 
the standard $\Lambda$CDM model.
We conclude that the lensing MFs are powerful probe of cosmology
only if mask effects are correctly taken into account.
\end{abstract}



\section{INTRODUCTION}

An array of recent observations such as
the cosmic microwave background (CMB) anisotropies \citep[e.g.][]{2011ApJS..192...18K,2013arXiv1303.5076P} 
and the large-scale structure \citep[e.g.][]{2006PhRvD..74l3507T,2010MNRAS.404...60R}
established the standard $\Lambda$CDM model.
The energy content of the present-day universe is
dominated by dark energy and dark matter, and
the primordial density fluctuations, which seeded all rich structure that we observe today,
were generated through inflation in the very early universe. 
A few important questions still remain such as the nature of dark energy,
the physical properties of dark matter, and the exact mechanism that generates
the primordial density fluctuations.

Gravitational lensing is a powerful method to study matter 
distribution \citep[e.g.][]{2012MNRAS.420.3213O}.
Future weak lensing surveys are aimed at measuring cosmic shear
over a wide area of more than a thousand square degrees. 
Such observational programmes include 
the Subaru Hyper Suprime-Cam (HSC) \footnotemark[1],  
the Dark Energy Survey (DES) \footnotemark[2], 
and the Large Synoptic Survey Telescope (LSST) \footnotemark[3].
\footnotetext[1]{\rm{http://www.naoj.org/Projects/HSC/j\_index.html}} 
\footnotetext[2]{\rm{http://www.darkenergysurvey.org/}}
\footnotetext[3]{\rm{http://www.lsst.org/lsst/}}
Space missions such as Euclid and WFIRST are also promising.
The large set of cosmic shear data will enable us to improve 
the constraints on cosmological parameters which will provide
important clues to the mysterious dark components.

A variety of statistics are proposed to characterize the
large-scale matter distribution.
Minkowski Functionals (MFs) are among the most useful statistics to extract 
non-Gaussian information from a two-dimensional or three-dimensional field.
For example, MFs have been applied to the observed CMB maps
and provided comparable constraints to those obtained using the CMB bispectrum 
\citep{2008MNRAS.389.1439H}.
\citet{2001ApJ...552L..89M} and \citet{2001ApJ...551L...5S}
studied $\Omega_{m}$-dependence of weak lensing MFs. 
More recently, \citet{2012PhRvD..85j3513K} showed that the lensing 
MFs contain significant cosmological information, beyond the
power-spectrum, whereas
\citet{2012ApJ...760...45S} showed that weak lensing MFs 
can be used to constrain
the statistical properties of the primordial density fluctuations.

These previous studies on weak lensing MFs often consider idealized cases.
However, many observational effects are present in real weak lensing measurements,
for example, imperfect shape measurement due to seeing and optical distortion, 
selection effects of galaxies, uncertain redshift
distribution of galaxies due to photometric redshift error
\citep[e.g.][]{2000A&amp;A...363..476B},
noise-rectification biases
\citep[e.g.][]{2000ApJ537555K,2001A&A366717E,2003MNRAS.343..459H},
and complicated survey geometry due to masked regions.
Some of these effects on cosmic shear power spectrum analysis 
have been already studied
\citep[e.g.][]{2006MNRAS.366..101H,2011MNRAS.412...65H}.
A comprehensive study of observational effects on lensing MFs
is also needed in order to fully exploit the data from 
upcoming wide cosmology surveys.

In the present paper, we study the impact of masked regions on the measurement of 
weak lensing MFs. Masking effect could be one of the major 
systematics 
because MFs are intrinsically morphological quantities.
We use a large set of numerical simulations to critically examine
the effect of masking.
We then directly measure the lensing MFs from real observational data 
obtained from a Subaru survey. We compare the observed MFs with the results of our
ray-tracing simulations that explicitly include the effect of masked regions.

The rest of the present paper is organized as follows.
In Section \ref{sec:MFs}, we summarize the basics of MFs and how to estimate 
MFs from observed shear field.
In Section \ref{sec:data}, we describe the data used in this paper and 
the details of numerical simulations 
of gravitational lensing.
In Section \ref{sec:res}, we show the results of the impact of masked regions on lensing MFs. 
We also perform a simple analysis to characterize the impact of masked regions on cosmological
constraints from lensing MFs.
We then represent the comparison with observed MFs and ray-tracing simulation results.
Concluding remarks and discussions are given in Section~\ref{sec:con}.

\section{MINKOWSKI FUNCTIONALS}
\label{sec:MFs}
\subsection{Basics}

MFs are morphological statistics for some smoothed random field 
above a certain threshold.
In general, for a given $D$-dimensional smoothed field $\mathbb{S}^{D}$,  
one can calculate $D+1$ MFs $V_{i}$.
On $\mathbb{S}^2$, one can thus define 2+1 MFs $V_{0}, V_{1}$, and $V_{2}$.
For a given threshold, $V_{0}$, $V_{1}$, and $V_{2}$ describe 
the fraction of area,
the total boundary length of contours, 
and the integral of the geodesic curvature $K$ along the contours,
respectively.
MFs are defined, for threshold $\nu$, as
\begin{eqnarray}
V_{0}(\nu) &\equiv& \frac{1}{4\pi}\int_{Q_{\nu}}\, {\rm d}S, \label{eq:V0_def} \\
V_{1}(\nu) &\equiv& \frac{1}{4\pi} \int_{\partial Q_{\nu}}\, \frac{1}{4} {\rm d}\ell , \label{eq:V1_def} \\
V_{2}(\nu) &\equiv& \frac{1}{4\pi} \int_{\partial Q_{\nu}}\, \frac{1}{2\pi}K{\rm d}\ell , \label{eq:V2_def}
\end{eqnarray}
where $Q_{\nu}$ and $\partial Q_{\nu}$ represent the excursion set 
and the boundary of the excursion set for a smoothed field $u(\bd{\theta})$.
They are given by
\beqa
Q_{\nu} = \{\bd{\theta}\, |\, u(\bd{\theta}) > \nu\}, \\
\partial Q_{\nu} =\{ \bd{\theta} \, |\, u(\bd{\theta}) = \nu \}.
\eeqa
For a two-dimensional Gaussian random field, one can calculate the expectation values 
for MFs analytically
\citep{1986PThPh..76..952T}:
\beqa
V_{0}(\nu)&=&\frac{1}{2}\left[1-{\rm erf}\left( \frac{\nu-\mu}{\sigma_{0}}\right)\right],
\label{eq:v0_gauss} \\
V_{1}(\nu)&=&\frac{1}{8\sqrt{2}}\frac{\sigma_1}{\sigma_0}\exp\left( -\frac{(\nu-\mu)^2}{\sigma_0^2}\right),
\label{eq:v1_gauss} \\
V_{2}(\nu)&=&\frac{\nu-\mu}{2(2\pi)^{3/2}}\frac{\sigma_{1}^2}{\sigma_{0}^3}\exp \left( -\frac{(\nu-\mu)^2}{\sigma_0^2} \right),
\label{eq:v2_gauss}
\eeqa
where $\mu=\langle u \rangle$, $\sigma_{0}^2 = \langle u^2 \rangle - \mu^2$, and
$\sigma_{1}^2 = \langle |\nabla u|^2 \rangle$.

\subsection{Estimation of Lensing MFs from Cosmic Shear Data}
\label{sec:est_MF}
We summarize how to estimate lensing MFs from observed shear data.
Let us first define the weak lensing mass maps that correspond to
the smoothed lensing convergence field $\kappa$:
\beqa
{\cal K} (\bd{\theta}) = \int {\rm d}^2 \phi \ \kappa(\bd{\theta}-\bd{\phi}) U(\bd{\phi}), \label{eq:ksm_u}
\eeqa
where $U$ is the filter function to be specified below.
We can calculate the same quantity by smoothing the shear field $\gamma$ as
\beqa
{\cal K} (\bd{\theta}) = \int {\rm d}^2 \phi \ \gamma_{t}(\bd{\phi}:\bd{\theta}) Q_{t}(\bd{\phi}), \label{eq:ksm}
\eeqa
where $\gamma_{t}$ is the tangential component of the shear at position $\bd{\phi}$ relative to 
point $\bd{\theta}$.
The filter function for the shear field $Q_{t}$ relates to $U$ by
\beqa
Q_{t}(\theta) = \int_{0}^{\theta} {\rm d}\theta^{\prime} \ \theta^{\prime} U(\theta^{\prime}) - U(\theta).
\eeqa
We consider $Q_{t}$ to be defined with a finite extent.
In this case, one finds 
\beqa
U(\theta) = 2\int_{\theta}^{\theta_{o}} {\rm d}\theta^{\prime} \ \frac{Q_{t}(\theta^{\prime})}{\theta^{\prime}} - Q_{t}(\theta),
\eeqa
where $\theta_{o}$ is the outer boundary of the filter function.

In the following, we consider the truncated Gaussian filter (for $U$) as
\beqa
U(\theta) &=& \frac{1}{\pi \theta_{G}^{2}} \exp \left( -\frac{\theta^2}{\theta_{G}^2} \right)
-\frac{1}{\pi \theta_{o}^2}\left( 1-\exp \left(-\frac{\theta_{o}^2}{\theta_{G}^2} \right) \right), \\
Q_{t}(\theta) &=& \frac{1}{\pi \theta^{2}}\left[ 1-\left(1+\frac{\theta^2}{\theta_{G}^2}\right)\exp\left(-\frac{\theta^2}{\theta_{G}^2}\right)\right],
\label{eq:filter_gamma}
\eeqa
for $\theta \leq \theta_{o}$ and $U = Q_{t} = 0$ elsewhere.
Throughout the present paper, we adopt $\theta_{G} = 1^{\prime}$ and $\theta_{o} = 15^{\prime}$.
Note that this choice of $\theta_{G}$ corresponds to an optimal smoothing scale 
for the detection of massive galaxy clusters
using weak lensing with $z_{\rm source}$ = 1.0 \citep{2004MNRAS.350..893H}.

We follow \citet{2012JCAP...01..048L}
in calculating the MFs from pixelated ${\cal K}$ maps.
We convert a 
weak lensing field ${\cal K}$ to $x = ({\cal K} - \langle {\cal K} \rangle)/\sigma_{0}$
where $\sigma_{0}$ is the standard deviation of ${\cal K}$.
In binning the thresholds, we set $\Delta x = 0.2$ from $x=-5$ to $x = 5$.
It is possible that the above normalization affects the MFs through
the variance of $\sigma_0$ for each field.
In light of this, \citet{1987ApJ...321....2W} suggest an alternative 
definition that a density contour with a certain threshold $\nu_{V}$ 
is related to the fraction of volume $f$ where  
\beqa
f = (2\pi)^{-1/2}\int_{\nu_{V}}^{\infty} e^{-t^2/2} {\rm d}t.
\label{eq:volume}
\eeqa
Using $\nu_{V}$ instead of $x$ apparently avoids the normalization issue 
for MFs at least technically.
However, even with $\nu_{V}$, 
we cannot eliminate the effect of the variance between multiple fields    
because the $f - \nu_{V}$ mapping needs to be done for each field or for each sample, 
rather than by using some global quantity calculated for all the samples.
We have explicitly tested the effect of the sample variance 
of $\sigma_{0}$ on MFs against $\nu_{V}$ and $x$
using 1000 Gaussian simulations
\footnote{We generate the Gaussian convergence maps for LCDM cosmology
using the fitting formula of \citet{2003MNRAS.341.1311S}
to calculate the matter power spectrum $P(k; z)$. We then 
integrate the matter power spectrum over redshift $z$,
convolved with a weighting function for the source redshift $z_{\rm source} =1$.
Each convergence map is defined on $2048^2$ grid points with 
an angular grid size of 0.15 arcmin.}.

\begin{figure}
\begin{center}
    \includegraphics[clip, width=1.\columnwidth]{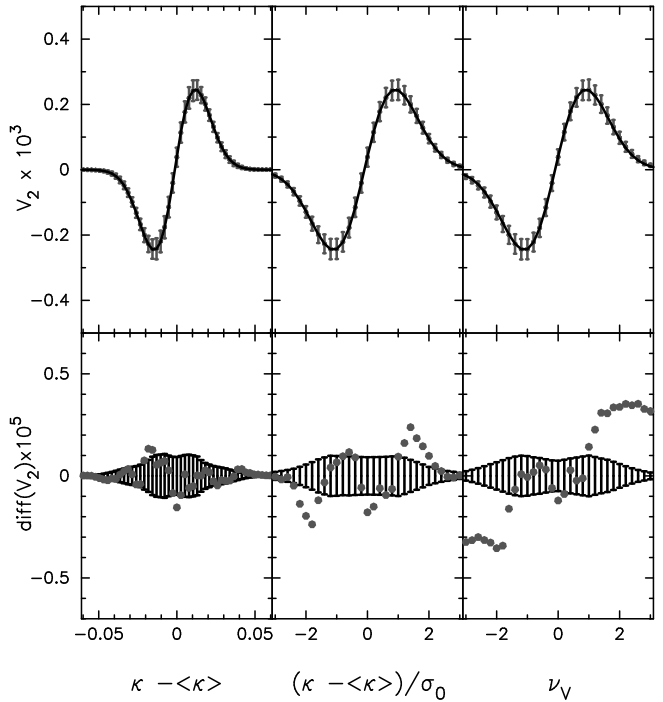}
    	\caption{
	The effect of sample variance of field variance $\sigma_{0}$ on MFs.
	We compare the mean $V_{2}$ over 1000 maps
	with the Gaussian prediction of Eq (\ref{eq:v2_gauss}).
	$V_2$ in the left panel is calculated without 
         normalization whereas that in the middle panel is
         calculated for each ${\cal K}$ field normalized by its variance
         and that in the right panel is calculated for each $\nu_{V}$
         (see text and the horizontal axes label).
        The gray points in the lower portion show the differences between the mean $V_{2}$ 
        and the Gaussian prediction. The differences are also compared with
        the variance of $V_{2}$ estimated from 
	our 1000 gaussian maps (black error bars), i.e. the standard deviation of $V_{2}$ 
        divided by $\sqrt{1000}$.	
	 \label{fig:comp_def_threshold}
	}
    \end{center}
  \end{figure}

In Figure \ref{fig:comp_def_threshold}, 
we compare the mean of $V_{2}$ MFs over our 1000 Gaussian maps 
with the Gaussian prediction given by Eq.~(\ref{eq:v2_gauss}).
For the Gaussian prediction, we calculate the quantities 
$\langle {\cal K} \rangle$, 
$\sigma_{0}$ and $\sigma_1$ by averaging over 1000 realizations;
these quantities serve as `global` values.
The error bars in each plot represent the variance of $V_{2}$ around 
the global mean.
The three panels differ in that the MFs are plotted as a function of,
from left to right, 
${\cal K} - \langle {\cal K} \rangle$,
$({\cal K} - \langle {\cal K} \rangle)/\sigma_0$,
and $\nu_{V}$, respectively.
The apparent variation of the MFs, namely error bars, in the middle 
and right panels 
is partly caused by the variance of
the measured $\sigma_0$ for each field.
In the lower panels, we show the difference between the mean $V_{2}$ and 
the Gaussian prediction. The difference should be compared with
the field variance that is indicated by error bars.
Note that the difference from the Gaussian prediction is larger
than the field variance when the MFs are evaluated 
with normalization as $({\cal K}-\langle {\cal K} \rangle)/\sigma_{0}$
or by using $\nu_{V}$ associated with volume fraction (Eq. (\ref{eq:volume})).
As expected, the Gaussian prediction describes
the mean MFs well as long as the MFs are evaluated $without$ normalization of 
${\cal K}$ by $\sigma_0$ (left panel).
However, we cannot use weak lensing field ${\cal K}$ directly 
when we compare theoretical predictions with the observation 
of a limited area (with masks). Theoretical predictions
for MFs are always given as a function of some normalized threshold.  
One thus needs either to de-normalize the theoretical prediction
by using an appropriate variance for the observed field, or
to normalize the observed ${\cal K}$ in some way.
In other words, field-to-field variance of the weak lensing MFs
is caused partly by the variance $\sigma_0$, and thus statistical
analysis such as cosmological parameter estimation should be
done by including the field variance of $\sigma_0$.
In the rest of this paper, we simply use the normalized field 
$x=({\cal K} - \langle {\cal K} \rangle)/\sigma_{0}$ for estimation of MFs.
When we estimate lensing MFs on a ${\cal K}$ map with mask, 
we discard the pixels within $2\theta_{G}$ from the mask boundaries,
because ${\cal K}$ data on such regions are affected by the lack of shear data. 

\section{DATA}
\label{sec:data}
\subsection{Suprime-Cam}

We use the $i^{\prime}$-band data 
from the Subaru/Suprime-Cam
data archive ${\sf SMOKA}$\footnotemark[1].
\footnotetext[1]{\rm{http://smoka.nao.ac.jp/}}
The observation is characterized as follows.
The area is contiguous with at least four pointings, 
the exposure time for each pointing is longer than 1800 sec, 
and the seeing full width at half-maximum (FWHM) is better than 0.65 arcsec.
The data are dubbed ``COSMOS" in Table A1 in \citet{2012MNRAS.425.2287H}.

We conservatively use the data only within a 15 arcmin radius from the field 
center of Suprime-Cam, because
the point spread function (PSF) 
becomes elongated significantly outside of the central area, 
which may make PSF correction inaccurate.
Then mosaic stacking is performed with ${\tt SCAMP}$ \citep{2006ASPC..351..112B} 
and ${\tt SWarp}$ \citep{2002ASPC..281..228B}. 
We use ${\tt SExtractor}$ \citep{1996A&AS..117..393B}
 and $hfindpeaks$ of the software ${\tt IMCAT}$ software \citep{1995ApJ...449..460K},
 and then the two catalogs are merged by matching positions of the detected 
objects with a tolerance of 1 arcsec.

For weak lensing analysis, we follow the KSB method 
\citep{1995ApJ...449..460K, 1997ApJ...475...20L, 1998ApJ...504..636H}.
Stars are selected in the standard way by identifying the appropriate branch in the magnitude
half-light radius ($rh$) plane, along with the detection significance cut $S/N > 10$. 
We found that the number density of stars is $\sim 1 \ {\rm arcmin}^{-2}$.
We use the galaxy images that satisfy the following three conditions;
(i) the detection significance of $S/N > 3$ and $\nu > 10$ 
where $\nu$ is an estimate of the peak significance given by $hfindpeaks$,
(ii) $rh$ is larger than the stellar branch,
and (iii) the AB magnitude is in the range of $22 < i^{\prime} < 25$ 
(where MAG\verb|_|AUTO given by ${\tt SExtractor}$ is used for the magnitude and 
slightly different from \citet{2012MNRAS.425.2287H}). 
The resulting number density of galaxies $n_{\rm gal}$ is then 15.8 ${\rm arcmin}^{-2}$. 
We measure the shapes of the objects using $getshapes$ in ${\tt IMCAT}$, 
and correct for the PSF using the KSB method. 
The $rms$ of the galaxy ellipticities after the PSF correction is 0.314.

Next, we define data and masked regions by using the observed positions of the 
source galaxies as follows. We map the observation area onto 
rectangular pixels of width 0.15 arcmin. For each pixel, we check
if there is a galaxy within $\theta_D = 0.4$ arcmin from the pixel center.
If there are no galaxies, then the pixel is marked as a mask pixel. 
After performing the procedure for all the pixels,
the marked pixels are masked regions, whereas the other pixels
are data regions. However, we unmask ``isolated'' masked pixels whose
surrounding pixels are all data pixels.

\begin{figure}
\begin{center}
    \includegraphics[clip, width=0.98\columnwidth]{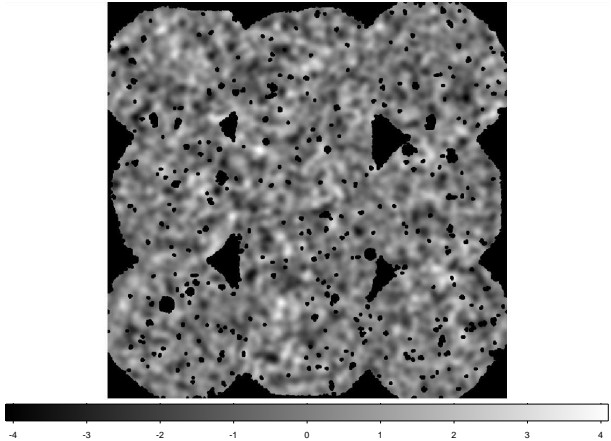}
    	\caption{
	 The reconstructed lensing field ${\cal K}$ from the Subaru 
         Suprime-Cam data.
	 The convergence ${\cal K}$ is computed from the ellipticity 
         of 102342 source galaxies 
         by Eq.(\ref{eq:ksm}).
	 The masked survey area (black portion) covers 0.34 ${\rm deg}^2$.
	 The grey-scale bar shows the value of 
	 $({\cal K} - \langle {\cal K}\rangle)/\sigma_0$.
	 \label{fig:obsmap}
	}
    \end{center}
  \end{figure}

Weak lensing convergence field ${\cal K}$ is computed from the galaxy 
ellipticity data as in Eq.~(\ref{eq:ksm})
on regular grids with a grid spacing of 0.15 arcmin.
The resulting mass map includes masked regions as shown in Figure \ref{fig:obsmap}.
The masked regions cover 0.34 ${\rm deg}^2$ in total. 
The unmasked regions are found to be 1.79 ${\rm deg}^2$. 
Note that we use only 0.575 ${\rm deg}^2$ in unmasked regions for lensing MFs analysis
because we remove the ill-defined pixels within $2\theta_G=2$ arcmin 
from the boundary of the mask.

\subsection{Ray-tracing Simulation}
\label{subsec:sim}

In order to study the impact of masked regions on lensing MFs,
we use 1000 weak gravitational lensing ray-tracing 
simulations from \citet{2009ApJ...701..945S}
\footnote{For the simulations, the adopted cosmology is consistent with 
WMAP3 results \citep{2007ApJS..170..377S}.}.
The ray-tracing simulations are performed on light-cone outputs 
that are generated by arranging multiple simulation boxes. 
Briefly, small- and large-volume $N$-body 
simulations are 
placed to cover a past light-cone of a hypothetical observer 
with an angular extent of $5^{\circ}\times 5^{\circ}$, from redshift 
$z=0$ to $z = 3.5$, similarly to the methods in  
\citet{2000ApJ...537....1W}
and 
\citet{2001MNRAS.327..169H}.
We set the source redshift $z_{\rm source} =1 $ for the ray-tracing simulations.
Each map is defined on $2048^2$ grid points with 
an angular grid size of 0.15 arcmin. 
Details of the ray-tracing simulations are found in \citet{2009ApJ...701..945S}.

It is well-known that the intrinsic ellipticities of source galaxies induce 
noises to lensing shear maps.
We model the noise by adding random ellipticities drawn from a
two-dimensional Gaussian to the simulated shear data. 
We set the root-mean-square of intrinsic ellipticities to be 0.314 
and the number of source galaxies is set 15.8 ${\rm arcmin}^{-2}$.
The values are obtained from the actual weak lensing observations 
described in Section 3.1.

\section{RESULT}
\label{sec:res}

\subsection{Masking Effect on Lensing MFs}

We first discuss the overall effect of masking on the lensing MFs.
To this end, we use ray-tracing simulations of weak gravitational 
lensing described in Section \ref{subsec:sim}.
We pay particular attention to non-Gaussian features in the case 
with masks. 
The total non-Gaussianity 
probed by the lensing MFs $\Delta V_{i}^{\rm obs}$ is given by
\beqa
\Delta V_{i}^{\rm obs} = V_{i}({\rm masked})-V_{i}^{G}({\rm masked}),
\label{eq:del_MFs_obs}
\eeqa
where $V_{i}(\rm masked)$ is $i$-th MF on a masked map and
$V_{i}^{G}(\rm masked)$ is the Gaussian term of $V_{i}(\rm masked)$.

\begin{figure*}
\begin{center}
    \includegraphics[clip, width=0.62\columnwidth]{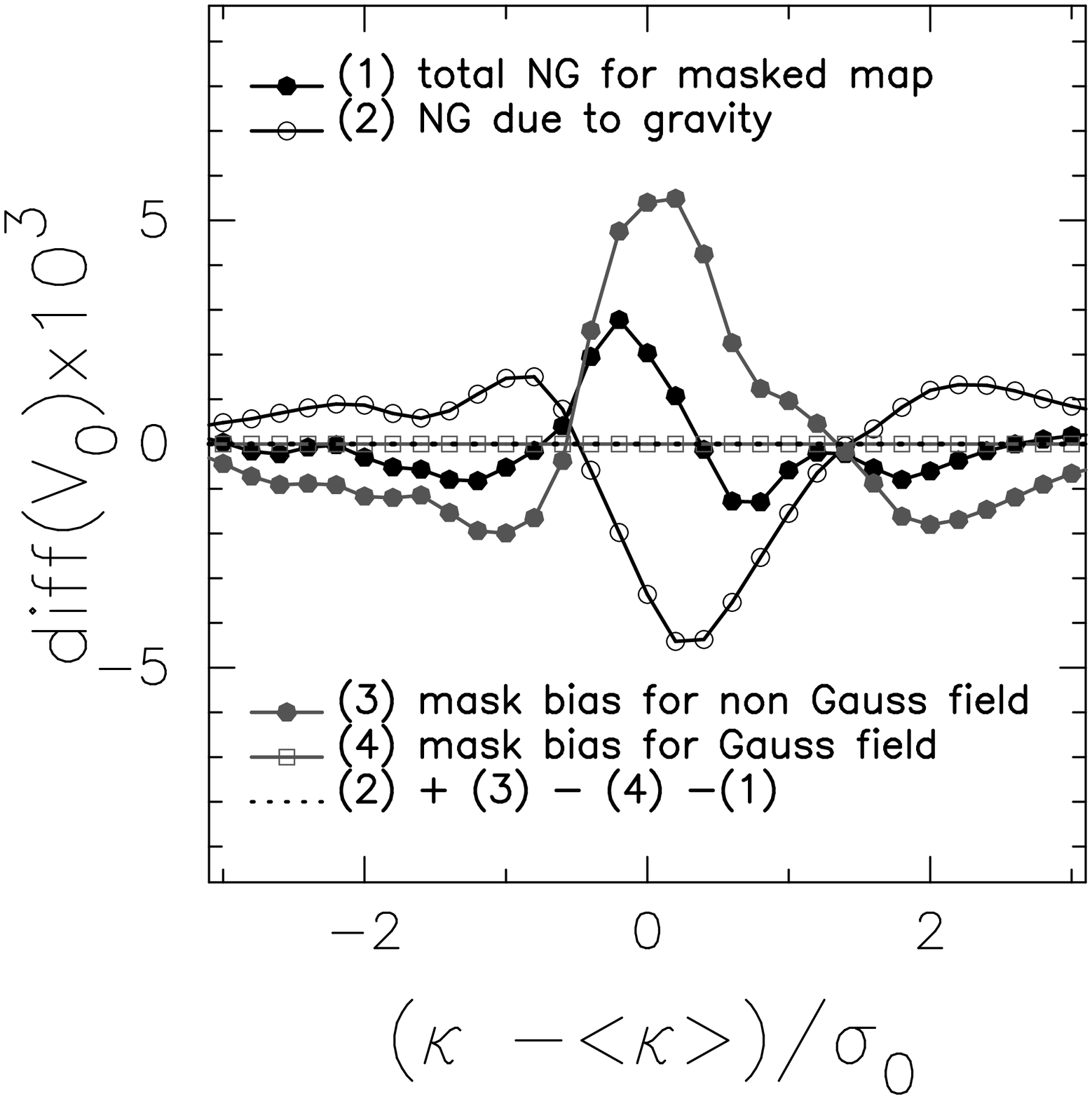}
    \includegraphics[clip, width=0.62\columnwidth]{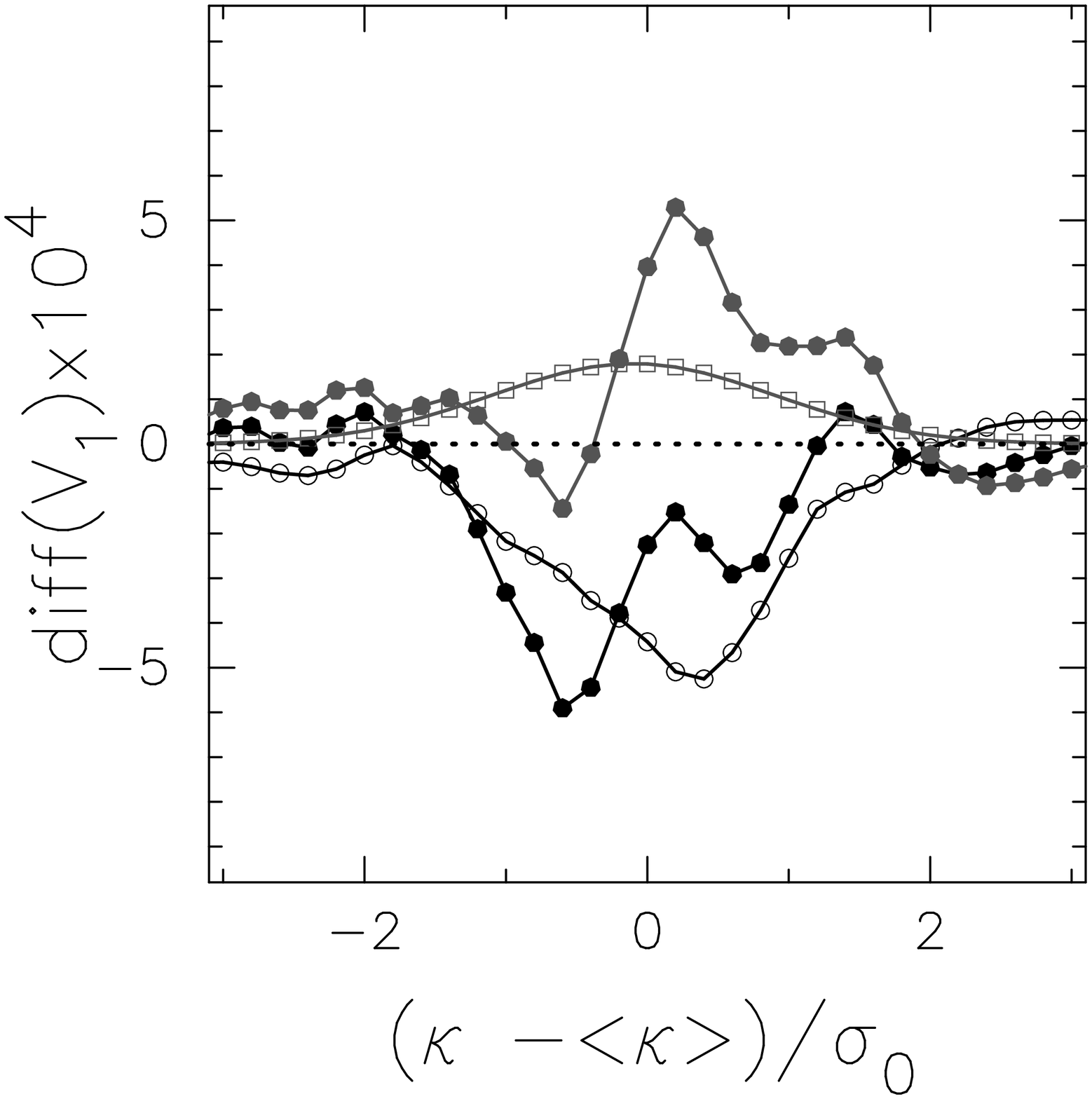}
    \includegraphics[clip, width=0.62\columnwidth]{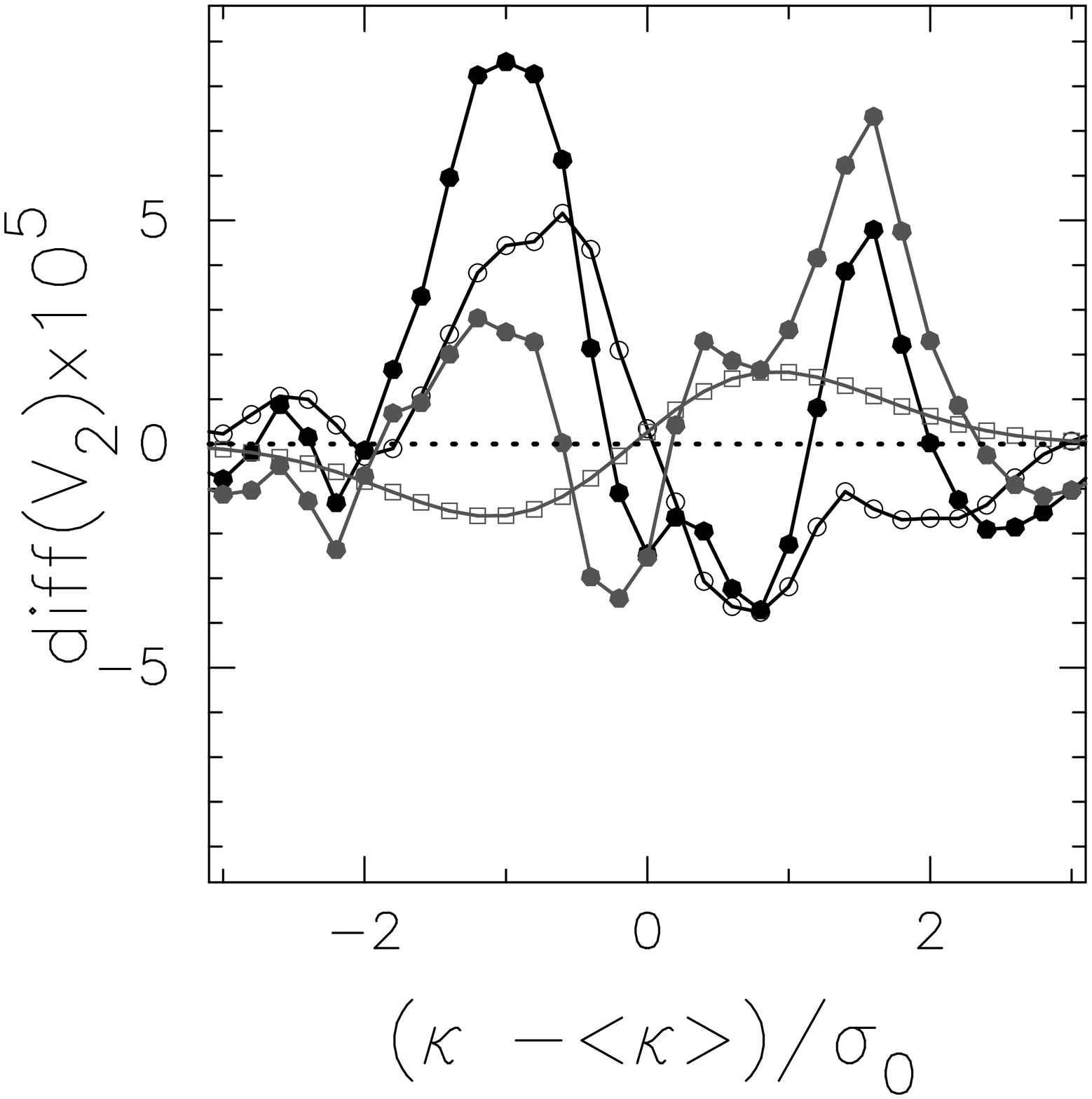}
    	\caption{
	We plot the differences between the lensing MFs on masked 
        ray-tracing simulation maps and the Gaussian term.
        The total non-Gaussianity obtained from the masked 
        maps $\Delta V_{i}^{\rm obs}$ 
        (black line with closed circle),
        the non-Gaussianity caused by non-linear gravitational 
	growth $\Delta V_{i}^{\rm gravity}$ 
	(black line with open circle), and 
	the bias of lensing MFs due to masked regions for ray-tracing maps
	$\Delta V_{i}^{\rm bias}$ (gray line with closed circle).
        We also plot the Gaussian term of $\Delta V_{i}^{\rm bias}$ 
        (gray line with open square).
	See the definition of each component given by 
	Eq. (\ref{eq:del_MFs_obs})-(\ref{eq:del_MFs_bias_gauss}).
	\label{fig:maskNG}
	}
    \end{center}
\end{figure*}

We can then decompose $\Delta V_{i}^{\rm obs}$ into three components:
\beqa
\Delta V_{i}^{\rm obs} &=& \Delta V_{i}^{\rm gravity} + \Delta V_{i}^{\rm bias} - \Delta V_{i}^{\rm bias, G}, \label{eq:del_MFs_obs3} \\
\Delta V_{i}^{\rm gravity} &=& V_{i}({\rm unmasked})-V_{i}^{G}({\rm unmasked}) 
\label{eq:del_MFs_grav}, \\
\Delta V_{i}^{\rm bias} &=& V_{i}({\rm masked}) - V_{i}({\rm unmasked}) 
\label{eq:del_MFs_bias}, \\
\Delta V_{i}^{\rm bias,G} &=& V_{i}^{G}({\rm masked}) - V_{i}^{G}({\rm unmasked}),
\label{eq:del_MFs_bias_gauss}
\eeqa
where $\Delta V_{i}^{\rm gravity}$ represents the non-Gaussianity induced by 
non linear gravitational growth,
$\Delta V_{i}^{\rm bias}$ describes the mask bias of MFs for non-Gaussian maps,
and
$\Delta V_{i}^{\rm bias, G}$ corresponds to the Gaussian term of $\Delta V_{i}^{\rm bias}$.
In order to calculate these quantities, we first need to calculate $V_{i}^{G}(\rm masked)$ 
and $V_{i}^{G}(\rm unmasked)$. For this purpose, we measure the following three 
quantities from 1000 masked ray-tracing maps:
\beqa
\mu=\langle {\cal K} \rangle, \ \sigma_{0}^2=\langle {\cal K}^2 \rangle -\mu^2, \
\sigma_{1}^2=\langle |\nabla {\cal K}|^2 \rangle.
\label{eq:mean_var}
\eeqa
The same quantities are measured also for the unmasked lensing maps. 
We can then estimate $V_{i}^{G}(\rm masked)$ and $V_{i}^{G}(\rm unmasked)$
using these quantities and Eq.~(\ref{eq:v0_gauss})-(\ref{eq:v2_gauss}). 
For the Gaussian terms, we also consider the correction of the finite 
binning effect pointed out by \citet{2012JCAP...01..048L}.
The correction is needed because the threshold $\nu$ 
to calculate
the MFs $V_{1}$ and $V_{2}$ is not continuous but discrete with some finite width.
We calculate the correction by integrating the analytic formula 
(Eq.~(\ref{eq:v1_gauss}),(\ref{eq:v2_gauss})) for finite binning width
(see \citet{2012JCAP...01..048L} for details).
We then estimate $V_{i}({\rm masked})$ and $V_{i}({\rm unmasked})$ 
directly from masked and unmasked maps.
Figure \ref{fig:maskNG} shows the various non-Gaussian contributions 
(Eq.~(\ref{eq:del_MFs_obs3})-(\ref{eq:del_MFs_bias_gauss})) 
calculated directly from 1000 masked ray-tracing maps.
We find that $\Delta V_{i}^{\rm bias}$ is comparable to $\Delta V_{i}^{\rm gravity}$ in the ray-tracing maps.
The mask bias $\Delta V_{i}^{\rm bias}$ contributes significantly to
the observed non-Gaussianity $\Delta V_{i}^{\rm obs}$. 
Note also that $\Delta V_{i}^{\rm bias,G}$ is sub-dominant although not negligible for $V_{1}$ and $V_2$.
Clearly the mask bias can be a significant contaminant for cosmological parameter estimation using the lensing MFs.
In the following, we include the bias effect when comparing simulation data and observations.

\subsection{Impact of Masking on Cosmological Parameter Estimation}
We next study cosmological information content in the lensing MFs with masks.
An important quantity is the cumulative signal-to-noise
ratio $S/N$ for lensing MFs, which is defined by
\beqa
(S/N)^{2}={\bd \mu}^{t}{\bd C}^{-1}{\bd \mu},
\eeqa
where ${\bd \mu}$ is a data vector that consists of 
the lensing MFs $V_{0}$, $V_{1}$, and $V_{2}$, and 
${\bd C}$ is the covariance matrix.
In order to calculate $(S/N)^2$, we construct the data vector from a set 
of lensing MFs as 
\beqa
\{\mu_{i} \}
=\{
V_{0}(x_1),...,V_{0}(x_{10}),
V_{1}(x_1),...,V_{1}(x_{10}),\nonumber \\
V_{2}(x_1),...,V_{2}(x_{10})
\},
\eeqa
where $x_{i}=({\cal K}_{i}-\langle {\cal K} \rangle)/\sigma_{0}$
is the binned normalized lensing field.
We calculate the covariance matrix of MFs using 1000 ray-tracing simulations.

\begin{figure}
\begin{center}
    \includegraphics[clip, width=0.9\columnwidth]{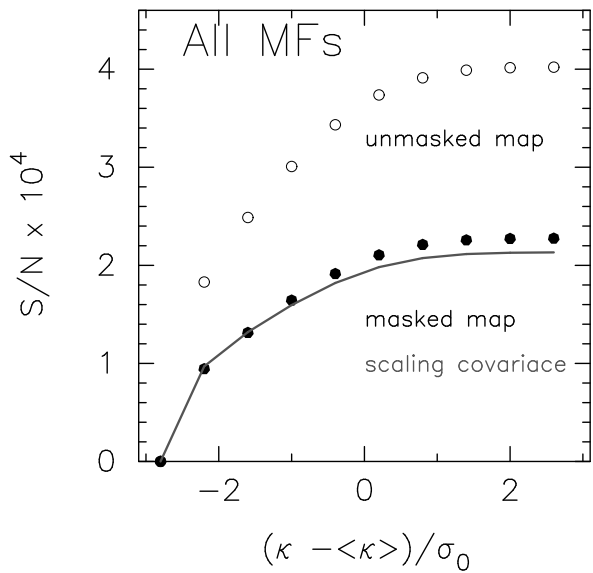}
    	\caption{
	The cumulative signal-to-noise ratio for the weak lensing MFs.
	The horizontal axis represents the maximum value of binned lensing field 
        used in the calculation of $S/N$.
	The open circles are the $S/N$ for unmasked 'clean' lensing maps 
        whereas the black points are
	for masked maps.
	The solid line shows $S/N$ obtained by scaling the covariance 
        	matrices of MFs with the effective survey area.
	We adopt the masked regions used for the Subaru Suprime-Cam data (see Figure \ref{fig:obsmap}).
	 \label{fig:allMFs_SN}
	}
    \end{center}
\end{figure}

Figure \ref{fig:allMFs_SN} 
shows the cumulative signal-to-noise ratio $S/N$ 
as a function of $x_{i}$.
Clearly the information content is reduced by a factor of two in the case with mask.
The degradation is explained by the reduced effective area.
The solid line shows $S/N$ by scaling ${\bd C}^{-1}$ with 
the effective survey area. It closely matches the 
$S/N$ calculated directly from the masked maps. 
For a Gaussian random field, we expect that the variance of MFs should be inversely 
proportional to the effective survey area
\citep[e.g.][]{1998NewA....3...75W,2009ApJ...691..547W}. 
The result shown in Figure \ref{fig:allMFs_SN} suggests that the effective survey 
area mainly determines how much cosmological information we can gain from weak lensing MFs.

Let us further quantify the overall impact of bias of the lensing MFs.
We perform the following simple analysis to investigate the effect of 
the mask bias on cosmological parameter estimation.
For each realization $r$ of our simulations, we calculate the $\chi^2$ value as follows,
\beqa
\chi^{2}(r) = (\mu_{i}(r)-\mu^{\rm theory}_{i}){\bd C}^{-1}
(\mu_{j}(r)-\mu^{\rm theory}_{j}),
\eeqa 
where $\mu_{i}(r)$ is the estimated lensing MFs from each realization $r$
and $\mu^{\rm theory}_{i}$ is the theoretical template for a given cosmology.
In practice, we assume that $\mu^{\rm theory}_{i}$ is the average over our 1000 
ray-tracing simulations with or without masks.
The lensing MFs $\mu_{i}(r)$ are estimated for each masked map,
and then we use the covariance matrices of the MFs obtained from a total of 1000 masked maps.
If $\mu_{i}(r)$ follows the Gaussian distribution,  
the distribution of $\chi^2(r)$ should follow a genuine $\chi^2$ distribution.
We can then clearly see the impact of bias due to masking
on cosmological constraints by comparing 
the resulting distribution of $\chi^2(r)$ 
for $\mu^{\rm theory}_{i}$ estimated from unmasked maps.

\begin{figure}
\begin{center}
    \includegraphics[clip, width=0.9\columnwidth]{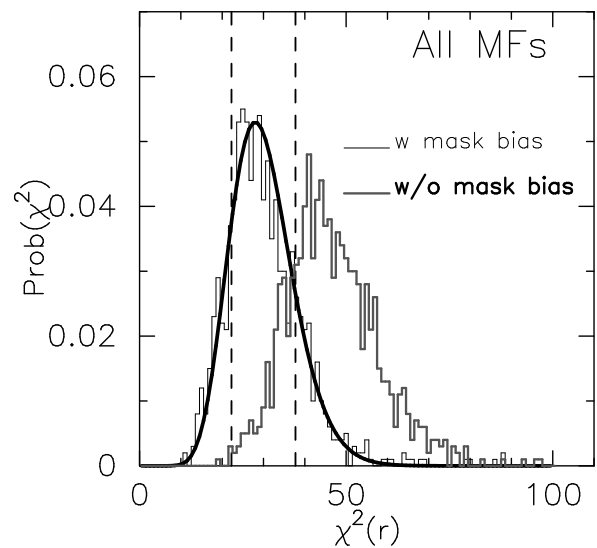}
    	\caption{
          We compare the distribution of $\chi^2(r)$ for
	 $\mu^{\rm theory}_{i}$ evaluated for 1000 masked maps (black histogram)
	 with that for $\mu^{\rm theory}_{i}$
	 evaluated for 1000 unmasked maps.
	 The thick solid lines is a genuine chi-square distribution with 30 degrees of freedom,
	 and dashed line represents the one sigma region.
	 \label{fig:test_mask_bias}
	}
    \end{center}
\end{figure}

Figure \ref{fig:test_mask_bias} shows the resulting distribution of $\chi^2(r)$ 
for our 1000 masked ray-tracing simulations.
The black histogram is the probability of $\chi^2(r)$ for the corresponding model 
using the average MFs 
over the masked maps whereas the gray one is for the unmasked maps.
The thick solid line is a genuine $\chi^2$ distribution with 30 degrees of freedom,
and the dashed line indicates the $1\sigma$ region for the $\chi^2$ values.
We find an excellent agreement between the thin histogram and the solid line.
This means that the binned lensing MFs $\mu_{i}(r)$ can be described well by a Gaussian distribution.
Interestingly, most of the resulting $\chi^2(r)$ without mask lie outside $1\sigma$ regions.
When we do not take account of bias due to masked regions,
55.3$\%$, 59.4$\%$, 74.9$\%$ and 85.4$\%$ of the realizations lies outside
$1\sigma$ regions of the $\chi^2$ values for $V_{0}$, $V_{1}$, $V_2$ and all MFs.
We conclude that the bias of lensing MFs due to masked regions can 
crucially affect a cosmological parameter estimation.

\subsection{Application to Subaru Suprime-Cam Data}

\begin{figure*}
\begin{center}
    \includegraphics[clip, width=0.62\columnwidth]{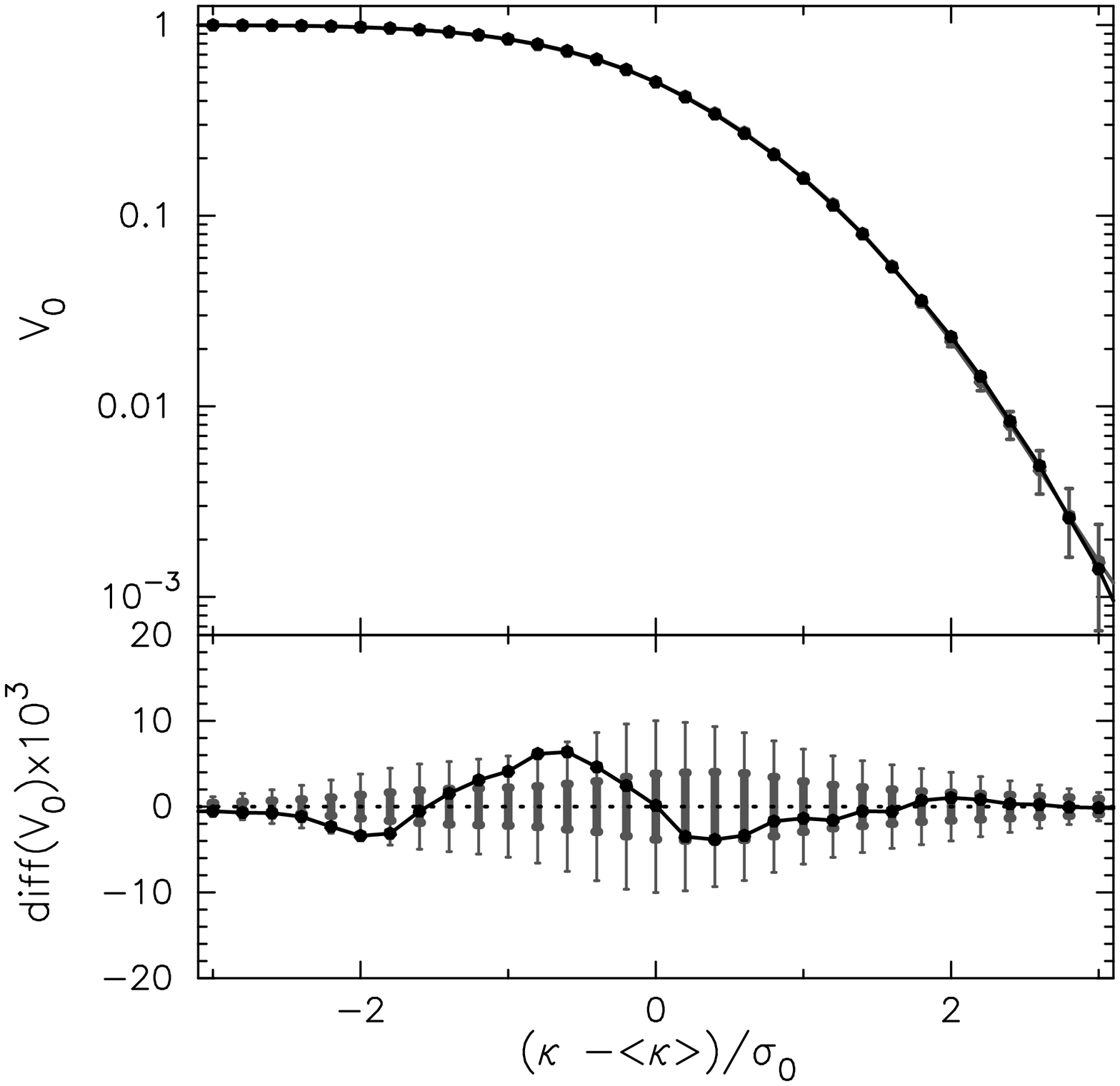}
    \includegraphics[clip, width=0.62\columnwidth]{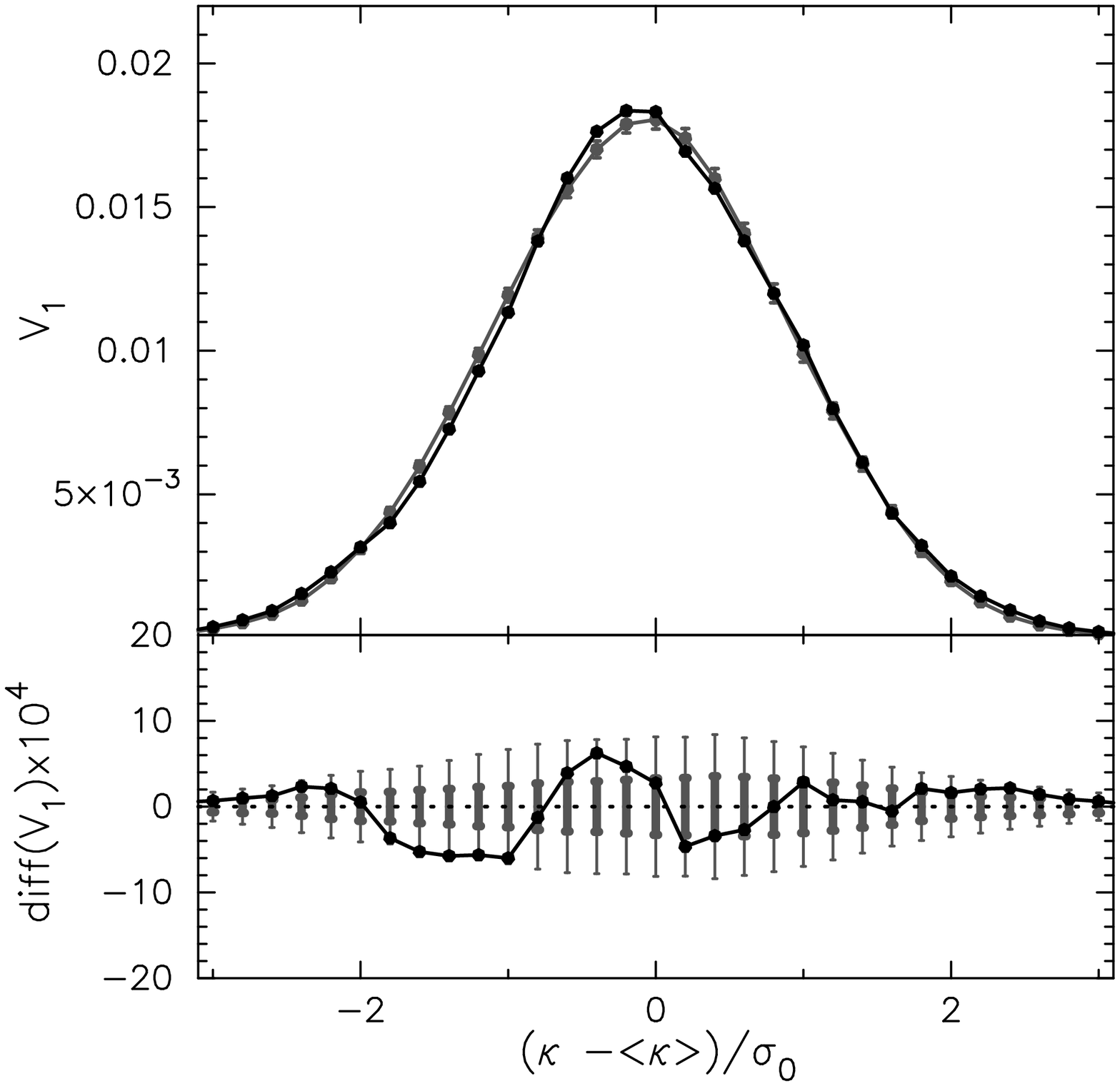}
    	\includegraphics[clip, width=0.62\columnwidth]{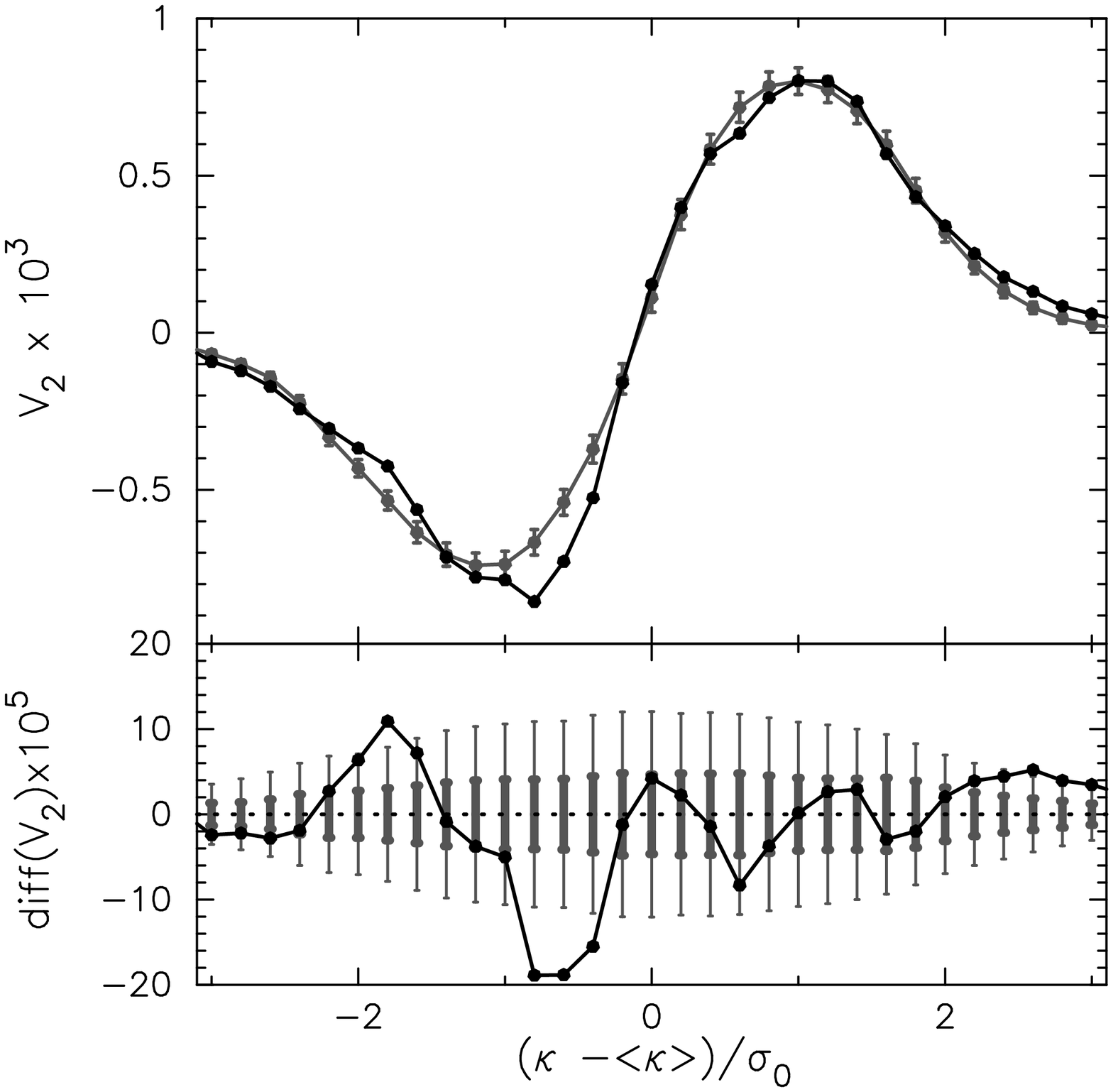}
    	\caption{
	 We compare the observed MFs with those from cosmological ray-tracing simulations.
	 In the upper panels, the black line shows the observed lensing MFs and the gray one 
         indicates the simulation results.
	The gray error bars show the cosmic variance obtained from 1000 ray-tracing simulations.
	In the lower panels, the black line shows the difference between the obtained MFs 
        and the simulation results.
	The thin error bars are the sum of the cosmic variance and the statistical error 
	while the thick error bars indicate only the cosmic variance. 
	The statistical errors are estimated from 1000 randomized galaxy catalogues.
	 \label{fig:obs_vs_RT}
	}
    \end{center}
  \end{figure*}

It is important to test whether we can extract cosmological information 
from masked noisy shear data using the lensing MFs. 
To this end, we use available Subaru Suprime-Cam data. 
We analyze the observed weak lensing map by using the statistics
derived from a large set of ray-tracing simulations. 
We include observational effects directly in our simulations, 
i.e., masked regions and shape noises as described in Section \ref{subsec:sim}.
Figure \ref{fig:obs_vs_RT} compares the lensing MFs for the Subaru data and 
those calculated for the ray-tracing simulations. We plot the MFs $V_0, V_1$, and $V_2$ in the top panels.
In the bottom panels, the thick error bars show the cosmic variance of lensing MFs estimated
from our 1000 simulated maps, whereas the thin error bars are the sum of the cosmic variance 
and the statistical error. We estimate the statistical error from 1000 randomized realizations, 
in which the ellipticity of each source galaxy is rotated randomly.
The statistical error is approximately $\sim 1.5$ times the cosmic variance for each bin.
In order to quantify the consistency of our results, we perform a so-called $\chi^2$ analysis.
We compute the $\chi^2$ statistics for the observed lensing MFs,
\beqa
\chi^2 = (d_{i}-m_{i}) {\bd C_{\rm cv+stat}}^{-1} (d_{j}-m_{j})
\eeqa
where $d_{i}$ is the lensing MFs in the $i$-th bin for observation, 
$m_{i}$ is the theoretical model,
and ${\bd C_{\rm cv+stat}}$ is the covariance matrix of lensing MFs including
the cosmic variance and the statistical error.
The cosmic variances are estimated from 1000 ray-tracing simulations, 
and the statistical errors are computed from
1000 randomized galaxy catalogs.
We estimate $m_{i}$ by averaging the MFs over 1000 ray-tracing simulations.
We use 10 bins in the range of $x=[-3,3]$ for each MF.
For the binning, we have a sufficient number of simulations to 
estimate the covariance matrix of the lensing MFs.
The resulting value of $\chi^2$ per number of freedoms
is $\chi^2/n_{\rm dof}=3.35/10, 9.69/10, 12.8/10$ and 29.6/30 
for $V_{0}$, $V_{1}$, $V_{2}$ and all the MFs.
The analysis includes the cosmic variance and the statistical
error as well as the mask effect.
We conclude that the observed lensing MFs are consistent with 
the standard $\Lambda$CDM cosmology.  

\section{SUMMARY AND CONCLUSION}
\label{sec:con}
We have used a large number of numerical simulations to examine how masked regions
affect the lensing MFs by adopting the actual sky-mask used for a Subaru observation.
We have then compared the observed lensing MFs 
with the results of cosmological simulations to address
whether the observed MFs are consistent with the standard cosmological model.

The weak lensing MFs are affected by the lack of cosmic shear data due mostly to 
foreground contamination.
We have used 1000 ray-tracing simulations with masked regions 
and with realistic shape noises, 
to show that the non-Gaussianities detected by the MFs do not solely come from
gravity induced non-Gaussianities. 
Masked regions significantly contaminate the $pure$ gravitational
signals. The bias is induced for the following two reasons:
(${\rm i}$) masked regions effectively reduce the number of 
sampling Fourier modes of cosmic shear
and (${\rm ii}$) masked regions introduce 
variance scatter of the reconstructed weak lensing mass field 
for each field of view.
The former can be corrected analytically at least for a Gaussian random field
as shown in the Appendix, while numerical simulations are 
needed to include the latter effect accurately.

We then perform a simple analysis to examine the impact of masked regions on
the cosmological parameter estimation.
From the cumulative signal-to-noise ratio for the lensing MFs, we have found that
the cosmological information content in the MFs can be largely determined by the effective survey area.
By studying the resulting distribution of the $\chi^2$ value for simulated maps with masks,
we can characterize how the ``mask bias'' of the MFs affects cosmological constraints.
We have shown that most of the resulting $\chi^2$ values are
found outside the expected one sigma region, when the mask is not considered.
Clearly the mask bias affect significantly the cosmological parameter estimation.

We have calculated the lensing MFs to the observed weak lensing shear map
obtained from a Subaru Suprime-Cam imaging survey.
Our analysis includes, in addition to the cosmic variance, 
the statistical error estimated from 1000 randomized galaxy catalogs.
The resulting $\chi^2/n_{\rm dof} = 29.6/30$ for all the MFs
suggests that the observed MFs are
consistent with the standard adopted $\Lambda$CDM cosmology.

Finally, we address the ability of the lensing MFs to constrain cosmological models.
By assuming a simple scaling of the covariance matrix of MFs by survey area,
we can reduce the error of MFs in each threshold bin by a factor of $\sim 20$ (100)
for upcoming weak lensing surveys with a 1000 (20000) ${\rm deg}^2$ survey area.
In the case of a 1000 ${\rm deg}^2$ survey, the error in each bin is 
translated to a $\sim 5\%$ difference in $\sigma_{8}$.
Similarly, for an LSST-like survey with a 20000 ${\rm deg^2}$ area,
the error in each bin is as small as $\sim0.5\%$ in $\sigma_{8}$.
The lensing MFs are a promising method for cosmology even with masked regions.
It is important to model the effect of mask accurately in order to
make the best use of lensing MFs for cosmological constraints.
The simplest way would be to use directly the observed mask on 
ray-tracing simulations as we have done in the present paper.
We will need such ray-tracing simulations covering a wide area of more than a thousand
square degrees for lensing MFs in upcoming wider surveys.

Further extensive studies are needed 
in order to devise a way to extract $pure$ cosmological information from the lensing MFs. 
It is important to study how much systematic non-Gaussianities are introduced by, for example, 
source galaxy clustering \citep[e.g.][]{1998A&A...338..375B},
source-lens clustering \citep[e.g.][]{2002MNRAS.330..365H},
the intrinsic alignment \citep[e.g.][]{2004PhRvD..70f3526H}, and 
inhomogeneous ellipticity noises due to inhomogeneous surface number density of sources
(M.Shirasaki et al., in preparation).
The upcoming wide-field surveys will provide highly-resolved lensing maps
but with complicated masked regions.
Our study in the present paper may be useful to properly analyze the data
and to accurately extract cosmological information from them.


\acknowledgments

We thank Chiaki Hikage and Masahiro Takada for useful discussions.
Masanori Sato provided us with their ray-tracing simulations data.
M.S. is supported by Research Fellowships of the Japan Society for 
the Promotion of Science (JSPS) for Young Scientists.
This work is supported by World Premier International Research Center 
Initiative (WPI Initiative), MEXT, Japan
and in part by Grant-in-Aid for Scientific Research from the JSPS Promotion of Science (23540324).
Numerical computations presented in this paper were in part carried out
on the general-purpose PC farm at Center for Computational Astrophysics,
CfCA, of National Astronomical Observatory of Japan.


\appendix

\section{EFFECT OF MASKS ON VARIANCE OF SMOOTHED CONVERGENCE FIELD}
\label{apdx2}
Here, we summarize the effect of masked regions on the variance of a smoothed convergence field
${\cal K}$.
When there are masked regions in a survey area, one needs to follow a special procedure in order to
construct a smoothed convergence field.
Let us define the masked region ${\cal M}_{s}(\bd{\theta})$ in a survey area as
\beqa
{\cal M}_{s}(\bd{\theta})=
\left \{
\begin{array}{ll}
1 &\ {\rm where} \ \bd{\theta} \ {\rm lies} \ {\rm in} \  {\rm data} \ {\rm region} \\
0 &\ {\rm otherwise}.
\end{array}
\right
.
\eeqa
When the area with mask ${\cal M}_{s}(\bd{\theta})$ 
is smoothed, there are ill-defined pixels due to 
the convolution between ${\cal M}_{s}$ and a filter function for smoothing $U$(\bd{\theta}).
We need to discard the ill-defined pixels to perform statistical analyses.
We therefore paste a new mask ${\cal M}_{1}(\bd{\theta})$ so that we can mask
the ill-defined pixels as well.
We then get
\beqa
{\cal K}^{\rm obs}(\bd{\theta})
&=&{\cal M}_{1}(\bd{\theta}){\cal K}_{1}(\bd{\theta}),
\eeqa
where
\beqa
{\cal K}_{1}(\bd{\theta})
&=&\int {\rm d}^2 \phi \ U(\bd{\theta}-\bd{\phi}){\cal M}_{s}(\bd{\phi})\kappa(\bd{\phi}).
\eeqa
The variance of the smoothed field is given by 
\beqa
\sigma_{0}^2 
&=&\frac{1}{S}\int {\rm d}^2 \theta \langle {\cal K}^{\rm obs}(\bd{\theta})^2 \rangle
\nonumber \\
&=&\frac{1}{S}\int {\rm d}^2 \theta {\cal M}_{1}(\bd{\theta}) \langle {\cal K}_{1}(\bd{\theta})^2 \rangle \nonumber \\
&=&\frac{1}{S}\int {\rm d}^2 \theta {\cal M}_{1}(\bd{\theta}) \int \frac{{\rm d}^2 \ell}{(2\pi)^2}\frac{{\rm d}^2 \ell^{\prime}}{(2\pi)^2} \langle {\cal K}_{1}(\bd{\ell}){\cal K}_{1}^{*}(\bd{\ell}^{\prime}) \rangle
\exp\left(i(\bd{\ell}-\bd{\ell}^{\prime}) \cdot \bd{\theta}\right) \nonumber \\
&=&\frac{1}{S}\int \frac{{\rm d}^2 \ell}{(2\pi)^2}\frac{{\rm d}^2 \ell^{\prime}}{(2\pi)^2}
{\cal M}_{1}(\bd{\ell}-\bd{\ell}^{\prime})
\langle {\cal K}_{1}(\bd{\ell}){\cal K}_{1}^{*}(\bd{\ell}^{\prime}) \rangle,
\eeqa
where we use the relation
${\cal M}_{1}(\bd{\theta})^2 = {\cal M}_{1}(\bd{\theta})$.
The Fourier mode of ${\cal K}_{1}$ is given by
\beqa
{\cal K}_{1}(\bd{\ell})
&=& \int {\rm d}^2 \theta {\rm d}^2 \phi \ 
U(\bd{\theta}-\bd{\phi}){\cal M}_{s}(\bd{\phi})\kappa(\bd{\phi})
\exp\left(i\bd{\ell} \cdot \bd{\theta}\right) 
\nonumber \\
&=&U(\bd{\ell})\int \frac{{\rm d}^2 \ell^{\prime}}{(2\pi)^2}
{\cal M}_{s}(\bd{\ell}^{\prime})\kappa(\bd{\ell}-\bd{\ell}^{\prime}), 
\nonumber \\
\langle{\cal K}_{1}(\bd{\ell}){\cal K}_{1}^{*}(\bd{\ell}^{\prime}) \rangle
&=&U(\bd{\ell})U^{*}(\bd{\ell}^{\prime})
\int \frac{{\rm d}^2 \ell_{1}}{(2\pi)^2}\frac{{\rm d}^2 \ell_{1}^{\prime}}{(2\pi)^2}
{\cal M}_{s}(\bd{\ell}_{1}){\cal M}_{s}^{*}(\bd{\ell}^{\prime}_{1})
\langle \kappa(\bd{\ell}-\bd{\ell}_{1}) \kappa^{*}(\bd{\ell}^{\prime}-\bd{\ell}^{\prime}_{1}) \rangle
\nonumber \\
&=& U(\bd{\ell})U^{*}(\bd{\ell}^{\prime})
\int \frac{{\rm d}^2 \ell_{1}}{(2\pi)^2}\frac{{\rm d}^2 \ell_{1}^{\prime}}{(2\pi)^2}
{\cal M}_{s}(\bd{\ell}_{s}){\cal M}_{s}^{*}(\bd{\ell}^{\prime}_{1})
(2\pi)^2 \delta^{(2)}(\bd{\ell}-\bd{\ell}_1-\bd{\ell}^{\prime}+\bd{\ell}_{1}^{\prime})
P_{\kappa}(|\bd{\ell}-\bd{\ell}_{1}|) 
\nonumber \\
&=&U(\bd{\ell})U^{*}(\bd{\ell}^{\prime})
\int \frac{{\rm d}^2 \ell_{1}}{(2\pi)^2}
{\cal M}_{s}(\bd{\ell}_{1}){\cal M}_{s}^{*}(\bd{\ell}_{1}+\bd{\ell}^{\prime}-\bd{\ell})
P_{\kappa}(|\bd{\ell}-\bd{\ell}_{1}|)
\eeqa

In the following, we assume that ${\cal M}_{1}(\bd{\theta})$ is large enough to cover
the ill-defined pixels due to smoothing (with a filter function) 
of the original masked region ${\cal M}_{s}(\bd{\theta})$.
This means that, with ${\cal M}_{1}$(\bd{\theta}), there remains
only $clean$ regions where
the smoothed convergence is not affected by the original masked regions 
${\cal M}_{s}(\bd{\theta})$.
In this case,
\beqa
{\cal K}^{\rm obs} (\bd{\theta}) \simeq {\cal M}_{1}(\bd{\theta}) \int {\rm d}^2 \phi \ U(\bd{\theta}-\bd{\phi}) \kappa(\bd{\phi}). \label{eq.app_kap_obs}
\eeqa
The fourier mode of ${\cal K}^{\rm obs}$ can then be given by
\beqa
{\cal K}^{\rm obs} (\bd{\ell}) 
&=& \int {\rm d}^2 \theta \ {\cal K}^{\rm obs} (\bd{\theta}) \exp\left(i \bd{\ell} \cdot \bd{\theta} \right)
\nonumber \\
&\simeq& \int {\rm d}^2 \theta \ {\cal M}_{1} (\bd{\theta}) \int {\rm d}^2 \phi \ U(\bd{\theta}-\bd{\phi}) \kappa(\bd{\phi}) \exp\left(i \bd{\ell} \cdot \bd{\theta} \right) 
\nonumber \\
&=& \int {\rm d}^2 \theta^{\prime} {\rm d}^2 \theta {\rm d}^2 \phi \ {\cal M}_{1}(\bd{\theta}^{\prime}) \delta^{(2)}(\bd{\theta}-\bd{\theta}^{\prime}) U(\bd{\theta}-\bd{\phi})
\kappa(\bd{\phi}) \exp\left(i \bd{\ell} \cdot \bd{\theta} \right) 
\nonumber \\
&=& \int \frac{{\rm d}^2 \ell^{\prime}}{(2\pi)^2} \int {\rm d}^2 \theta^{\prime} {\rm d}^2 \theta {\rm d}^2 \phi \ {\cal M}_{1}(\bd{\theta}^{\prime}) 
U(\bd{\theta}-\bd{\phi}) \kappa(\bd{\phi}) \exp\left(i \bd{\ell} \cdot \bd{\theta} \right) \exp\left( -i \bd{\ell}^{\prime} \cdot (\bd{\theta}-\bd{\theta}^{\prime})\right) 
\nonumber \\
&=& \int \frac{{\rm d}^2 \ell^{\prime}}{(2\pi)^2} {\cal M}_{1}(\bd{\ell}^{\prime})
\int {\rm d}^2 \theta {\rm d}^2\phi \ U(\bd{\theta}-\bd{\phi}) \kappa(\bd{\phi})
\exp\left( i(\bd{\ell}-\bd{\ell}^{\prime})\cdot\bd{\theta}\right) \nonumber \\
&=& \int \frac{{\rm d}^2 \ell^{\prime}}{(2\pi)^2} {\cal M}_{1}(\bd{\ell}^{\prime})
U(\bd{\ell}-\bd{\ell}^{\prime}) \kappa(\bd{\ell}-\bd{\ell}^{\prime}).
\eeqa
The variance of the smoothed convergence field is calculated as
\beqa
\sigma_{0}^{2}
&=&\frac{1}{S}\int {\rm d}^2 \theta \langle {\cal K}^{\rm obs}(\bd{\theta})^2 \rangle
\nonumber \\
&=&\frac{1}{S}\int {\rm d}^2 \theta 
\int \frac{{\rm d}^2 \ell}{(2\pi)^2}\frac{{\rm d}^2 \ell^{\prime}}{(2\pi)^2}
\langle {\cal K}^{\rm obs}(\bd{\ell}) {\cal K}^{\rm obs}(\bd{\ell}^{\prime})\rangle
\exp\left(-i(\bd{\ell}-\bd{\ell}^{\prime})\cdot \bd{\theta}\right)
\nonumber \\
&=&\frac{1}{S}\int \frac{{\rm d}^2 \ell}{(2\pi)^2} \ 
\langle {\cal K}^{\rm obs}(\bd{\ell}) ({\cal K}^{\rm obs})^{*}(\bd{\ell}) \rangle,
\eeqa
where the ensemble average of the Fourier mode is 
\beqa
\langle {\cal K}^{\rm obs}(\bd{\ell}) ({\cal K}^{\rm obs})^{*}(\bd{\ell}^{\prime}) \rangle
&\simeq& \int \frac{{\rm d}^2 \ell_{1}}{(2\pi)^2}\frac{{\rm d}^2 \ell_{1}^{\prime}}{(2\pi)^2}
{\cal M}_{1}(\bd{\ell}_{1}){\cal M}_{1}^{*}(\bd{\ell}^{\prime}_{1})
U(\bd{\ell}-\bd{\ell}_{1}) U^{*}(\bd{\ell}^{\prime}-\bd{\ell}^{\prime}_{1}) \nonumber \\
&& \hspace{50pt} \times \langle \kappa(\bd{\ell}-\bd{\ell}_{1})\kappa^{*}(\bd{\ell}^{\prime}-\bd{\ell}^{\prime}_{1})\rangle \nonumber \\
&=& \int \frac{{\rm d}^2 \ell_{1}}{(2\pi)^2}\frac{{\rm d}^2 \ell_{1}^{\prime}}{(2\pi)^2}
{\cal M}_{1}(\bd{\ell}_{1}){\cal M}_{1}^{*}(\bd{\ell}^{\prime}_{1})
U(\bd{\ell}-\bd{\ell}_{1}) U^{*}(\bd{\ell}^{\prime}-\bd{\ell}^{\prime}_{1}) \nonumber \\
&& \hspace{50pt} \times (2\pi)^{2}P_{\kappa}(|\bd{\ell}-\bd{\ell}_{1}|) \delta^{(2)}(\bd{\ell}-\bd{\ell}_{1}-\bd{\ell}^{\prime}+\bd{\ell}^{\prime}_1) \nonumber \\
&=& \int \frac{{\rm d}^2 \ell_{1}}{(2\pi)^2}
{\cal M}_{1}(\bd{\ell}_{1}){\cal M}_{1}^{*}(\bd{\ell}_{1}+\bd{\ell}^{\prime}-\bd{\ell})
|U(\bd{\ell}-\bd{\ell}_{1})|^2 P_{\kappa}(|\bd{\ell}-\bd{\ell}_{1}|).
\eeqa

\begin{figure}
\begin{center}
    \includegraphics[clip, width=0.45\columnwidth]{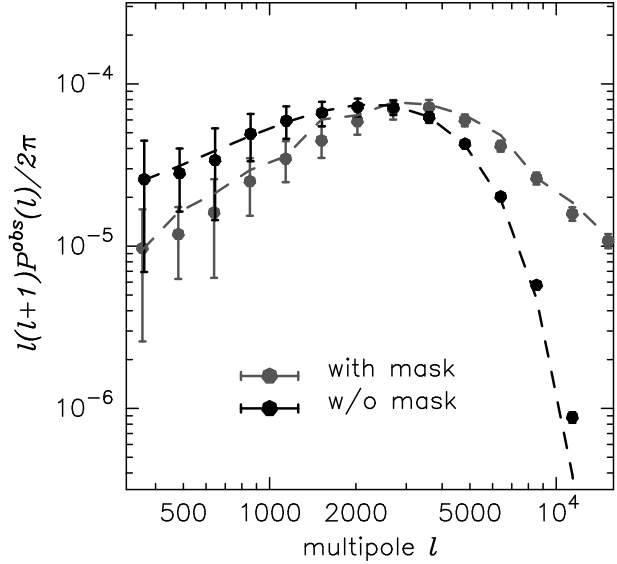}
    	\caption{
	We test the validity of Eq.(\ref{eq.app_kap_obs}).
	The gray points with error bars show $P^{\rm obs}(\ell)$ obtained from 1000 Gaussian maps with
	mask ${\cal M}_{s}(\bd{\theta})$.
	The gray dashed line is the theoretical prediction of Eq.(\ref{eq.app_Pobs}).
	The black points with error bars show $P^{\rm obs}(\ell)$ obtained 
        from 1000 maps without mask.
	The black dashed line is the input power spectrum smoothed by the Gaussian filter $U$. 
	 \label{fig:compare_mask_power}
	}
    \end{center}
  \end{figure}

We have checked the validity of Eq.(\ref{eq.app_kap_obs}) by using 1000 Gaussian simulations.
They are the same set of simulations as in Section \ref{sec:est_MF}.
For each Gaussian simulation, we paste the observed masked region ${\cal M}_{s}(\bd{\theta})$
from the Subaru Suprime-Cam observation.
The map is then smoothed with a Gaussian filter of Eq.(\ref{eq:ksm_u}).
The adopted smoothing scale is 1 arcmin.
In order to avoid the ill-defined pixels, we paste a new mask ${\cal M}_{1}(\bd{\theta})$,
which is constructed conservatively to cover the regions within two times the smoothing scale 
from the boundary of the original mask ${\cal M}_{s}(\bd{\theta})$.
We then calculate 
\beqa
P^{\rm obs}(\ell)
&\equiv& \langle {\cal K}^{\rm obs}(\bd{\ell}) ({\cal K}^{\rm obs})^{*}(\bd{\ell}) \rangle/S.
\eeqa
If ${\cal K}^{\rm obs}$ can be well-approximated by Eq.~(\ref{eq.app_kap_obs}),
this quantity should be given by
\beqa
P^{\rm obs}(\ell)
&\simeq& \frac{1}{S}\int \frac{{\rm d}^2 \ell_{1}}{(2\pi)^2}
|{\cal M}_{1}(\bd{\ell}_{1})|^2
|U(\bd{\ell}-\bd{\ell}_{1})|^2 P_{\kappa}(|\bd{\ell}-\bd{\ell}_{1}|).\label{eq.app_Pobs}
\eeqa
Figure \ref{fig:compare_mask_power} compares Eq.~(\ref{eq.app_kap_obs})
and Eq.~(\ref{eq.app_Pobs}).
Clearly Eq.~(\ref{eq.app_kap_obs}) is an excellent approximation
for the observed survey geometry.
The ill-defined pixels are efficiently masked by 
${\cal M}_{1}(\bd{\theta})$.
We also find the variance $\sigma_{0}^2$ decreases 
by a factor of $O(5\%)$.
This causes the bias of MFs even if the lensing field is Gaussian.

\bibliography{ref}

\end{document}